\documentclass[aps,nofootinbib,showpacs,twocolumn]{revtex4-1}
\usepackage[CJKbookmarks, pdftex, bookmarksnumbered, bookmarksopen, colorlinks, citecolor=blue, linkcolor=blue]{hyperref}
\usepackage[english]{babel}
\usepackage{amstext,amsmath,amssymb,amsfonts,bbm}
\usepackage[latin1]{inputenc}
\usepackage{graphicx}
\usepackage{epstopdf}
\usepackage{amsthm}
\usepackage{tocvsec2}
\usepackage{enumerate}
\usepackage{subfigure}
\usepackage{color}
\usepackage{tikz}

\begin{document}

\title{Statistical mechanics of covariant systems with multi-fingered time}

\author{Goffredo Chirco}
\affiliation{Max Planck Institute for Gravitational Physics, Albert Einstein Institute, Am M\"{u}hlenberg 1, 14476, Potsdam, Germany.}
\author{Thibaut Josset}
\affiliation{Aix Marseille Univ, Univ Toulon, CNRS, CPT, Marseille, France.}

\date{\today}

\begin{abstract} 
\noindent 
Recently, in [\textit{Class. Quantum Grav.} \textbf{33} (2016) 045005], the authors proposed a new approach extending the framework of statistical mechanics to reparametrization-invariant systems with no additional gauges. In this work, the  approach is generalized to systems defined by more than one Hamiltonian constraints (multi-fingered time). We show how well known features as the Ehrenfest-Tolman effect and the J\"{u}ttner distribution for the relativistic gas can be consistently recovered from a covariant approach in the multi-fingered framework. Eventually, the crucial role played by the interaction in the definition of a global notion of equilibrium is discussed.
\end{abstract}

\maketitle

\section{Introduction}

Thermodynamics and general relativity are well understood essential pillars of modern physics. Individually, these theories provides an extremely accurate description of nature. However, when considered together, their interplay raises unexpected questions: Why are the laws of black holes mechanics equivalent to the laws of thermodynamics \cite{Bardeen1973The-four-laws-o}? Why is Hawking radiation thermal  \cite{Hawking1974Black-hole-expl}? What are the degrees of freedom contributing to black hole entropy \cite{Bekenstein1973Black-Holes-and}? Is the Einstein equation an equation of state \cite{Jacobson1995Thermodynamics-}? Is gravity an entropic force \cite{Verlinde2011On-the-origin-o}?

Despite much effort, the nature of the interplay between thermodynamics and general relativity remains mostly obscure. In particular, no fully (general) relativistic statistical mechanics --taking spacetime's degrees of freedom into account-- has yet been found. One of the main obstacles in this sense is due to the central role played by \emph{time} in statistical mechanics. General relativity has forced us to reconsider the basic notions used in mechanics such as space, time, observable, reference frame. In particular, in a general covariant framework, the notion of time is considerably weakened\footnote{The ``problem of time'' plays a central role for several distinct issues \cite{Kuchar1992Time-and-Interp, Rovelli1995Analysis-of-the}, from the choice of an internal time variable in quantum gravity, to the emergence of a directional global time in cosmology.}. This necessarily leads to a serious modification of the notion of thermodynamic equilibrium. 

In \cite{Chirco2016Statistical-mec} the authors discussed a \emph{relational} reformulation of the notions of statistical state and equilibrium, for systems characterised by reparametrization invariance with no extra gauge symmetry. Such systems, defined by a vanishing Hamiltonian, have one-dimensional orbits, with no preferred parametrization. In other words, the time parameter appearing in the equations of motion is unphysical. In this context, the notion of statistical state was shown to rely on a suitable split of the system into two parts, one on which measurements are performed, and one used as a clock. From a relational perspective, the ideal Newtonian time of classical mechanics turned out to be just a peculiar example of ``clock system''.

In this paper, we move a step forward along the same line, by considering the case of reparametrization-invariant systems with additional gauge symmetries. Such systems are defined  by more than one vanishing constraint, so the orbits are multi-dimensional, their parametrization being still unphysical. For instance, a field theory like general relativity has one Hamiltonian constraint per point in space, and thus as many (unphysical) time parameters. Such systems are typically referred to as \emph{multi-fingered time} systems. 

We investigate the statistical mechanics of systems with multi-fingered time, making use of the formal approach defined in \cite{Chirco2016Statistical-mec} on single constraint systems. More precisely, we will argue for the crucial role of \emph{interaction} in the construction of the statistical mechanics of these systems. 

The paper is organized as follows. Section \ref{section_RelativisticSystems} reviews the Hamiltonian formulation of relativistic mechanics and the statistical mechanics deduced from it, in the case of a single constraint. Section \ref{section_TolmanEffect} probes a simple system with a two-fingered time, and derive the Ehrenfest-Tolman effect. Section \ref{section_Gas} recovers the statistical mechanics of a gas of special relativistic particles, starting from a completely covariant mechanical description. Finally, section \ref{section_Discussion} summarizes our results and briefly discusses the possible implications for gravity.

\section{Relativistic systems}\label{section_RelativisticSystems}

Classical mechanics provides a very good description of the evolution of observables in Newtonian time, but it becomes inappropriate for relativistic systems. Therefore, generalized Hamiltonian mechanics has been developed by Dirac \cite{Dirac1950Generalized-ham}. Following \cite{Rovelli2004Quantum-gravity}, we give here a brief summary of the formalism of generalized Hamiltonian mechanics, which will then be used throughout the paper.

\subsection{Mechanics}\label{subsection_RelativisticMechanics}

Let us consider a relativistic system described in a $n$-dimensional \emph{configuration space} $\mathcal{C}$, coordinatized by \emph{partial observables}  $(q^\alpha)$, whose dynamics is given by a reparametrization-invariant action $\int L(q^\alpha,\dot q^\alpha)d\tau$. Any such singular Lagrangian leads to a vanishing Hamiltonian, relating the different conjugate momenta $p_\alpha = \frac{\partial L}{\partial \dot q^\alpha}$. If there are extra gauge symmetries, one gets a set of \emph{constraints} $\{C^i=0~,~ 1 \leq i\leq k\}$. We call \emph{extended phase space} the $2n$-dimensional symplectic space $X=T^*\mathcal{C}=\{(q^\alpha,p_\alpha)\}$, equipped with $\omega_X = dp_\alpha \wedge dq^\alpha$, and \emph{presymplectic surface} the $2n-k$ subspace $\Sigma$ defined by the constraints.

\begin{figure}[h]
	\includegraphics[scale=0.3]{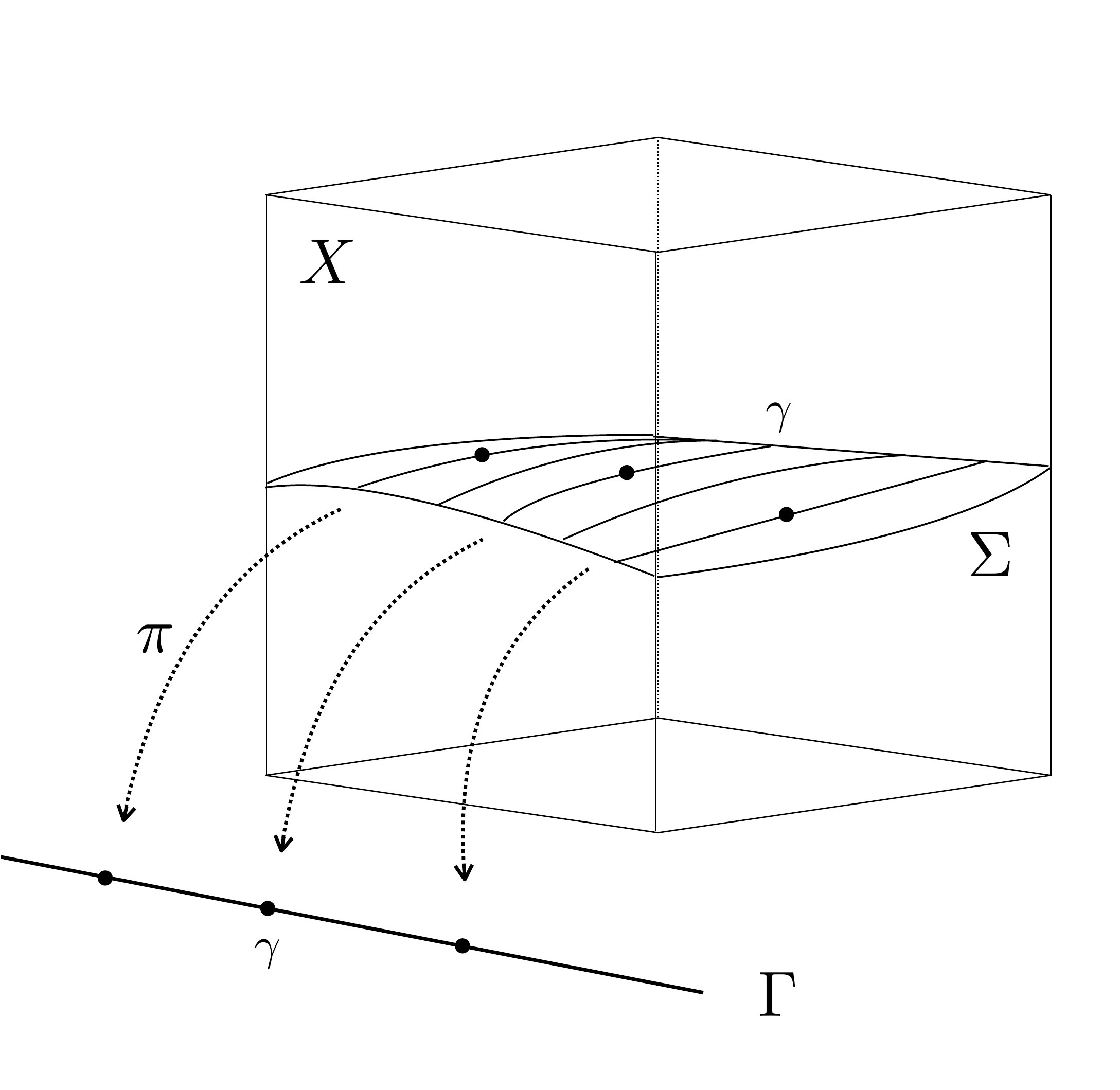}
	\caption{Generalized Hamiltonian mechanics.}
	\label{FigureRelativisticMechanics}
\end{figure}

The dynamics can be recast in the action $\int \left( p_\alpha \dot q^\alpha -N_i C^i(q^\alpha,p_\alpha) \right) d\tau$, where $\{N_i\}$ are Lagrange multipliers. The orbits are then given by Hamilton equations
\begin{equation}
	\frac{dq^\alpha}{d\tau} = N_i(\tau) \frac{\partial C^i}{\partial p_\alpha}~,~\frac{dp_\alpha}{d\tau} = -N_i(\tau) \frac{\partial C^i}{\partial q^\alpha}.
\end{equation}
Different choices for the functions $\{N_i\}$ corresponds to gauge-equivalent representations of the same $k$-dimensional motion. Finally, the \emph{physical phase space} $\Gamma$ is simply the set of motions, and it has dimension $2n-2k$ (see figure \ref{FigureRelativisticMechanics}).

\subsection{Statistical mechanics of reparametrization-invariant systems}\label{subsection_RelativisticStatisticalMechanics}

	The possibility of doing statistical mechanics of reparametrization-invariant systems, without additional gauge ($k=1$), has been recently discussed in \cite{Chirco2016Statistical-mec}. This new approach is based on a formulation of the ergodic hypothesis, without any reference to Newtonian time. In particular, it shows that, for a system made of two non-interacting subsystems
\begin{equation}\label{eq_split2}
	\begin{array}{c}
	X=X^a\times X^b ~,~ \omega_X = \omega_{X^a} + \omega_{X^b} , \\
	\text{and}~ C= C^a+C^b=0~,
	\end{array}
\end{equation}
the microcanonical ensemble for one subsystem (say $\mathcal{S}^b$), using the other one (here $\mathcal{S}^a$) as a clock, is well defined and simply reads\footnote{If $C^b$ is not the only conserved quantity for $\mathcal{S}^b$, the statistical state should include additional ``$\delta$-functions''.}
\begin{equation}\label{microcanonical}
	\frac{\delta(C^b-I)\, d\mu_{X^b}}{\int \delta(C^b-I)\, d\mu_{X^b}},
\end{equation} 
where $d\mu_{X^b}$ is the Liouville measure on $\left(X^b,\omega_{X^b}\right)$ and $I$ indicates the value of the function $C^b$ (preserved by the dynamics). It is important to notice that \eqref{microcanonical} does not depend on any specific choice of clock in $\mathcal{S}^a$, nor of its dynamics (given by $C^a$). Given a well defined notion of microcanonical ensemble, it is then possible to apply the standard tools of statistical mechanics. More precisely, if the subsystem of interest in turn splits into weakly interacting components
\begin{equation}\label{eq_split3}
	\begin{array}{c}
	 X=X^a\times X^b \times X^c ~,~ \omega_X = \omega_{X^a} + \omega_{X^b} + \omega_{X^c} ,  \\
	 \text{and} ~ C= C^a+C^b+C^c+V^{bc}=0 ,
	\end{array}
\end{equation}
one can generalize the notions of entropy and temperature, respectively
\begin{equation*}
	S^\alpha(I^\alpha) \underset{def}{=} \log \int \delta(C^\alpha- I^\alpha) d\mu_{X^\alpha} ~,~ \frac{1}{T_I^\alpha} \underset{def}{=}  \frac{dS^\alpha}{dI^\alpha} .
\end{equation*}
The equilibrium between $\mathcal{S}^b$ and $\mathcal{S}^c$, with respect to $\mathcal{S}^a$, is then simply given by the (generalized) zeroth law
\begin{equation}\label{eq_zerothlaw}
	T_I^b = T_I^c .
\end{equation}

	In brief, the existence of absolute Newtonian time and energy can be reformulated in a relational way, by means of a suitable \emph{split} of the whole system into subsystems. Doing so naturally extends statistical mechanics and thermodynamics to reparametrization-invariant systems (cf. \cite{Chirco2016Statistical-mec} for more details). Let us explore how this approach could be extended to systems with more than constraints.

\section{Ehrenfest-Tolman effect}\label{section_TolmanEffect}

In 1930, Tolman jointly applied the mass-energy equivalence (special relativity) and the equivalence principle (general relativity) to heat \cite{Tolman1930On-the-Weight-o}. He found that, for a fluid at equilibrium in a non-uniform gravitational field, the (proper) temperature is \emph{not} constant\footnote{This fact was actually glimpsed by Einstein \cite{Einstein1912Zur-theorie-des}, even before General Relativity was completed!}. In a second paper \cite{Tolman1930Temperature-Equ}, Tolman and Ehrenfest gave the following characterization of thermodynamical equilibrium, in a static spacetime:
\begin{equation}\label{eq_tolman}
T(\vec x) \sqrt{-g_{00}(\vec x)} = cste ,
\end{equation}
where $T$ is the temperature measured by a local observer at rest and $\sqrt{-g_{00}}$ is the norm of the timelike Killing vector field in stationary coordinates. Equation \eqref{eq_tolman} is now called Ehrenfest-Tolman effect, and has been re-derived in various manners \cite{Balazs1965On-thermodynami, Ebert1973Carnot-cycles-i, Rovelli2011Thermal-time-an, Haggard2013Death-and-resur}.

Despite the \emph{apparent} conflict between \eqref{eq_zerothlaw} and \eqref{eq_tolman}, we will see that our formalism, when applied to multi-fingered time systems, naturally leads to Ehrenfest-Tolman effect.

\subsection{Two non-relativistic systems}

Let us consider two non-relativistic systems $\mathcal{S}^a$ and $\mathcal{S}^b$ in some small regions\footnote{So that the local gravitational field is approximately uniform.} of a stationary spacetime. Let us denote  $\tau^a$ and $\tau^b$ the (local) proper times. Assuming for the time being that $\mathcal{S}^a$ and $\mathcal{S}^b$ do not interact, the global system is described in the extended phase space 
\begin{equation*}
	\begin{array}{l}
	X=X^a \times X^b = \{\tau^a,p_{\tau^a}, {\bold{q}^a}, {\bold{p}_a}, \tau^b,p_{\tau^b}, {\bold{q}^b}, {\bold{p}_b}\}, \\
	\omega = dp_{\tau^a} \wedge d\tau^a + d{\bold{p}_a} \wedge d{\bold{q}^a} + dp_{\tau^b} \wedge d\tau^b + d{\bold{p}_b} \wedge d{\bold{q}^b},
	\end{array}
\end{equation*}
by the two constraints
\begin{equation}\label{eq_FreeConstraintsTolman}
	\left\{ \begin{array}{lclcr} C^{a} &=& p_{\tau^a} + H^{a}({\bold{q}^a},{\bold{p}_a}) &=& 0 \\ 
						C^{b} &=& p_{\tau^b} + H^{b}({\bold{q}^b},{\bold{p}_b}) &=&  0 \end{array} \right. .
\end{equation}

Denoting $n^a+1$ (resp. $n^b+1$) the dimension of the configuration space of $\mathcal{S}^a$ (resp. $\mathcal{S}^b$), the orbits of the global system are two-dimensional surfaces, and the physical phase space $\Gamma$ has dimension $2(n^a+n^b)$, as expected from Newtonian mechanics.

\subsection{Gauge choices}

It may seem unusual to describe the dynamics of a non-relativistic system by the mean of two time observables. Let us pick a single variable $t$: without loss of generality, $\tau^{a} = t$ and $\tau^{b} = f(t)$. The symplectic form simply reads
\begin{equation*}
	\tilde \omega = d(p_{\tau^a}+f'(t) p_{\tau^b}) \wedge dt + d{\bold{p}_a} \wedge d{\bold{q}^a}  + d{\bold{p}_b} \wedge d{\bold{q}^b} .
\end{equation*}
Denoting $p_t=p_{\tau^a}+f'(t) p_{\tau^b}$ the conjugate momentum to $t$, the reduced system is then described in the extended phase space 
\begin{equation*}
	\tilde X = \{t,p_t,{\bold{q}^a},{\bold{p}_a},{\bold{q}^b},{\bold{p}_b}\} ,
\end{equation*}
by the single constraint
\begin{equation}\label{eq_tolmansingleconstraint}
	\tilde C = p_t + H^{a} ({\bold{q}^a},{\bold{p}_a}) + f'(t) H^{b}({\bold{q}^b},{\bold{p}_b}) = 0 .
\end{equation}

From the point of view of mechanics, any such gauge choice is leading to a well defined constrained system. In particular, the $2(n^a+n^b)$-dimensional physical phase space remains unchanged, no information has been lost. On the other hand, the affine gauge choice 
\begin{equation}\label{eq_affinegaugechoice}
	f(t) = \alpha t + \beta
\end{equation} is the only one leading to a time-independent Hamiltonian in \eqref{eq_tolmansingleconstraint}. In relational terms, it is the only choice leading to a split of the system between a (Newtonian) clock component, and the other partial observables. Thus, it is preferred from a \emph{statistical} mechanics perspective, to satisfy requirement \eqref{eq_split2}. However, the rate $\alpha$ is still arbitrary. In other words, time and energy scales can still be chosen independently for the two systems $\mathcal{S}^a$ and $\mathcal{S}^b$, thus it would be meaningless to compare their temperatures.

In the next paragraph, we illustrate how the \emph{interaction} between $\mathcal{S}^a$ and $\mathcal{S}^b$ selects a preferred gauge choice.

\subsection{Interaction and statistical state}

 Let us consider the modified constraint
\begin{equation*}
	\left\{ \begin{array}{lclcr} C^{a} &=& p_{\tau^a} + H^{a}({\bold{q}^a},{\bold{p}_a}) +V({\bold{q}^a},{\bold{q}^b},\tau^a,\tau^b)&=& 0 \\ 
						C^{b} &=& p_{\tau^b} + H^{b}({\bold{q}^b},{\bold{p}_b}) &=&  0 \end{array} \right. .
\end{equation*}

In Newtonian space-time, it would be natural to assume that the (instantaneous) interaction $V$ depends on the time variables only through $\tau^a-\tau^b$. In a stationary spacetime, $V$ depends only on $\frac{\tau^a}{\sqrt{-g_{00}^a}}-\frac{\tau^b}{\sqrt{-g_{00}^b}}$, where $g_{00}^a$ (resp. $g_{00}^b$) is the gravitational field seen by the subsystem $\mathcal{S}^a$ (resp. $\mathcal{S}^b$), in stationary coordinates. After gauge fixing, the single constraint \eqref{eq_tolmansingleconstraint} splits if and only if $\alpha = {\sqrt{-g_{00}^b}}~/{\sqrt{-g_{00}^a}}$, and in that case
\begin{equation*}
	\tilde C = p_t + H^{a}({\bold{q}^a},{\bold{p}_a}) + \frac{\sqrt{-g_{00}^b}}{\sqrt{-g_{00}^a}} H^{b}({\bold{q}^b},{\bold{p}_b}) + V({\bold{q}^a},{\bold{q}^b})= 0 .
\end{equation*}

\begin{figure}[h]
	\includegraphics[scale=0.8]{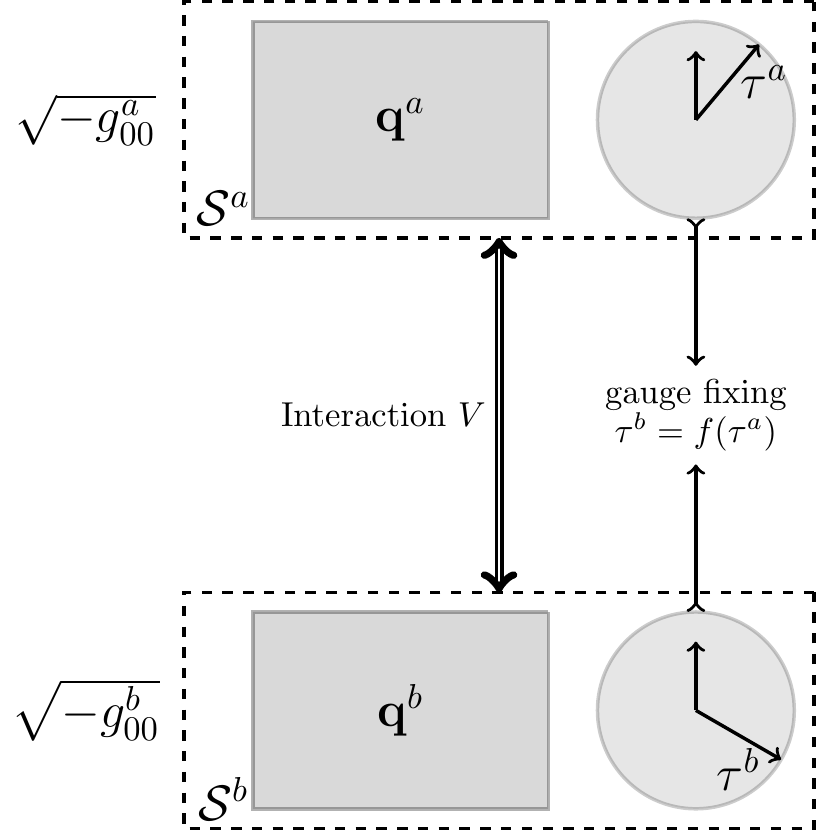}
	\caption{Two non-relativistic systems $\mathcal{S}^a$ and $\mathcal{S}^b$, evolving in a stationary, locally uniform, gravitational field, can be described either with their own (local) time observables (gauge redundancy) or using a single time variable. The interaction $V$ selects a preferred gauge, leading to an equilibrium state which satisfies Ehrenfest-Tolman effect.}
	\label{FigureTolmanEffect}
\end{figure}

Thus, by requiring condition \eqref{eq_split3} to be satisfied, the interaction selects a preferred gauge (see figure \ref{FigureTolmanEffect}). One can now write the zeroth law \eqref{eq_zerothlaw} in terms of generalized temperatures, and finally deduce Ehrenfest-Tolman relation
\begin{equation*}
	T^a\sqrt{-g_{00}^a}=  T^b\sqrt{-g_{00}^b}
\end{equation*}
for the proper temperatures.

	To summarize, it is clear that there is no way to define a temperature for the global system without considering the interaction. Ehrenfest-Tolman effect is due to the change in the weight of energy that is \emph{moved} in space, from one subsystem to the other one.

\section{Relativistic gas}\label{section_Gas}

After a long controversy about the correct statistical mechanics and thermodynamics in special relativity, the generalization of Maxwell distribution initially proposed by Jüttner in 1911 \cite{Juttner1911Das-Maxwellsche},
\begin{equation*}
	\gamma(\vec v)^{5} e^{-m \gamma(\vec v) / k_B T} ,
\end{equation*}
where $\gamma(\vec v) = \frac{1}{\sqrt{1-\vec v^2}}$, seems to be confirmed by recent numerical simulations \cite{Cubero2007Thermal-Equilib}. Using the momentum $\vec p =\nolinebreak m \gamma(\vec v) \vec v$, the probability distribution reads
\begin{equation}\label{eq_Juttnerdistribution}
	\frac{1}{Z} e^{-\beta^\mu p_\mu} d^3 \vec p d^3 \vec x ,
\end{equation}
where $\beta^\mu$ is a time-like 4-vector, reducing to $\left( \frac{1}{k_B T},0,0,0 \right)$ in the center of mass frame.

\subsection{Free special relativistic particles}

The generalized Hamiltonian formalism presented in section \ref{section_RelativisticSystems} is very convenient to keep manifest the symmetries of special relativistic systems. In particular, a free particle of mass $m$ is characterized by $p_\mu p^\mu + m^2 = 0$. A gas of non-interacting $N$ such particles is simply described in the $8N$-dimensional extended phase space
\begin{equation*}
	X = \underset{\alpha}{\times} \{ x^{(\alpha)\mu},p^{(\alpha)}_\mu\},
\end{equation*}
by the set of constraints
\begin{equation}\label{eq_FreeConstraintsGas}
	C^{(\alpha)} = p^{(\alpha)}_\mu p^{(\alpha)\mu}+m^2=0 ,\quad 1\leq \alpha \leq N.
\end{equation}
Motions are $N$-dimensional surfaces, and the physical phase space has dimension $6N$.

\subsection{Gauge choices}

Let us choose a single time-like variable $t$, related to the coordinates of each particle by
\begin{equation*}
	t = f_\alpha \left( x^{(\alpha)} \right),\quad 1\leq \alpha \leq N.
\end{equation*}

As in section \ref{section_TolmanEffect}, a generic gauge choice would lead to a time-dependent Hamiltonian (i.e. a constraint that does not split), and requirement \eqref{eq_split2} for statistical mechanics would not be satisfied.

As long as there is no interaction, it is possible to choose different Lorentz frames for different particles: $t=x'^{(\alpha)0}$, where $x'^{(\alpha)} = \Lambda(\alpha) x^{(\alpha)}$. In that case, the single constraint is
\begin{equation}\label{eq_relativisticgassingleconstraint}
	\tilde{C}=p_t + \sum_\alpha \sqrt{\left( \vec p'^{(\alpha)}\right)^2+m^2}=0 ,
\end{equation}
where $p'^{(\alpha)}_i = {\left[\Lambda(\alpha) \right]_i}^\mu p^{(\alpha)}_\mu$ is the conjugate momentum to $x'^{(\alpha)i}$. The constraint \eqref{eq_relativisticgassingleconstraint} splits into a clock part and a time-independent Hamiltonian. However, as the Lorentz transformations $\{\Lambda(\alpha)\}$ can be chosen arbitrarily, the temperature of the gas is ill-defined.

\subsection{Gas of weakly interacting particles}

Though, the picture changes when a coupling between the particles is turned on. Let us consider a (scattering) interaction term\footnote{$V$ can be thought as an effective potential coming from a (Lorentz invariant) interaction with an (relativistic) ambient field.}
 of the form $V\left[ \left(x^{(\alpha)\mu}-x^{(\beta)\mu} \right) \left(x_\mu^{(\alpha)}-x_\mu^{(\beta)} \right) \right]$. It is clear that the interaction part of the Hamiltonian constraint is time-independent only if one chooses the same Lorentz frame for all the particles. So, without loss of generality, let us take $\Lambda(\alpha)=\mathbb{I}$. Then, \eqref{eq_relativisticgassingleconstraint} reduces to the classical formulation of special relativistic particles:
\begin{equation*}
	\tilde{C} = p_t +\sum_\alpha \sqrt{\left(\vec p^{(\alpha)}\right)^2 + m^2} + \sum_{\alpha \neq \beta} V\left[ \left\|\vec x(\alpha) - \vec x(\beta)\right\| \right]=0 ,
\end{equation*}
where $V$ insures thermalization of the gas, but remains small compared to the kinetic energy. For any value of $p_t=-E$ (and total $3$-momentum\footnote{In the absence of external potential, the total $4$-momentum is conserved, insuring Lorentz covariance of the statistical state.} $\vec P$), one can write the microcanonical state \eqref{microcanonical}, and deduce Jüttner distribution \eqref{eq_Juttnerdistribution}.

The equilibrium statistical state is not selected by the Lorentz invariance of each free constraint \eqref{eq_FreeConstraintsGas}, but rather by the Lorentz invariance of the \emph{interaction} term $V$.

\section{Discussion}\label{section_Discussion}

	It is known since Boltzmann's work that a weak, mixing, interaction is necessary for reaching equilibrium. However, in general, the details of the interaction do not matter much, and its analysis is often relegated to the ergodic \emph{hypothesis}. In the case of relativistic systems described by several constraints, the role of the interaction is dramatically enhanced, as it entails whether the constraint can be split or not. Therefore, the interaction determines the subsystems for which it is meaningful to talk about equilibrium, and the corresponding clock systems. This led us to recover well known features (Ehrenfest-Tolman effect and J\"uttner distribution) from a covariant approach.

	The interplay between interaction and gauge has been studied in \cite{Rovelli2014Why-Gauge}, and its implications for statistical mechanics were suggested in the context of thermal time hypothesis\cite{Chirco2013Coupling-and-th}. In this work, we have taken a step further by showing how equilibrium states are selected in the whole set of possible statistical states.

	Trying to apply this formalism to general relativity, one faces a technical problem: how to find a nice gauge and identify a split of the system. This issue is very similar to the problem of time in quantum gravity. However, in the context of statistical mechanics one may relax requirement \eqref{eq_split3} into
\begin{equation*}
	C= C^a+C^b+C^c+V^{bc}+V^{abc}=0 ,
\end{equation*}
where $V^{abc} \ll V^{bc}$. In other words, the only physically reasonable assumption is that the different components are ``thermalizing'' faster than exchanging ``energy'' with the clock system.

\section{Acknowledgments}

We are grateful to Carlo Rovelli for careful reading of the draft and useful comments.
\\


\end{document}